\documentclass[preprint,aps,showpacs,preprintnumbers,amsmath,amssymb,nofootinbib]
{revtex4}
\usepackage{graphicx}
\usepackage{epsfig}
\begin{document}



\newcommand{\be}{\begin{equation}}
\newcommand{\ee}{\end{equation}}
\newcommand{\bea}{\begin{eqnarray}}
\newcommand{\eea}{\end{eqnarray}}
\newcommand{\nn}{\nonumber}
\def\CP{{\it CP}~}
\def\cp{{\it CP}}
\title{Perturbation to  TBM mixing and its phenomenological implications }

\author{Sruthilaya M.$^a$, Srinu Gollu$^{a,b}$ }
\affiliation{$^a$School of Physics, University of Hyderabad, Hyderabad-500 046, India \\
$^b$Government Degree College, Kuppam-517 425, Andhra Pradesh, India }

\begin{abstract}
To accommodate the recently observed  non-zero reactor mixing angle $\theta_{13}$, we consider the  
lepton mixing matrix as Tri-bimaximal mixing (TBM) form in the leading order along with a perturbation in neutrino sector. 
The perturbation is taken to be a rotation in 23 plane followed by a rotation in 13 plane, i.e., $R_{23}(\theta_{23}')R_{13}(\theta_{13}',\phi)$. 
We obtain the allowed values of the parameters $\theta_{23}'$, $\theta_{13}'$ and $\phi$, which can accommodate all the observed mixing angles 
consistently  and calculate the phenomenological observables such as the  Dirac CP violating phase ($\delta_{CP}$), 
Jarlskog invariant ($J_{CP}$), effective majorana mass $M_{ee}^{\nu}$, and $m_{\nu_e}$, the electron neutrino mass. 
We find that $\delta_{CP}$ can take any values between $0$ and $-\pi/2$ and $M_{ee}^{\nu}$ always comes below its experimental upper limit.

\end{abstract}

\pacs{14.60.Pq, 14.60.Lm}
\maketitle
\section{Introduction}
Discovery of neutrino oscillation confirmed that neutrinos have non-zero masses. It indicates at least one of the mass eigenstates is non-degenerate and 
the standard model neutrino flavour states are mixture of  mass eigenstates, $\nu_1$, $\nu_2$, and $\nu_3$, i.e.,
\begin{equation}
\left(
\begin{array}{ccc}
\nu_e \\
\nu_{\mu}\\
\nu_{\tau}
\end{array}
\right)=
\left(
\begin{array}{ccc}
U_{e1}&U_{e2}&U_{e3} \\

U_{\mu 1}&U_{\mu 2}&U_{\mu 3}\\
U_{\tau 1}&U_{\tau 2}&U_{\tau 3}
\end{array}
\right)
\left(
\begin{array}{ccc}
\nu_1 \\
\nu_2\\
\nu_3
\end{array}
\right),
\end{equation} 
where $U$ is the lepton mixing matrix known as PMNS matrix \cite{a11,a11a}, which can be parametrized in terms of three mixing angles and one 
CP violating phase ($\delta_{CP}$), if neutrinos are Dirac particles. There will be two more phases known as Majorana phases in addition to $\delta_{CP}$,
 if neutrinos are Majorana type.
In the standard parametrization,  PMNS matrix is represented as
\begin{equation}
V_{PMNS} = U_{PMNS}.P_\nu
 = \left( \begin{array}{ccc} c^{}_{12} c^{}_{13} & s^{}_{12}
c^{}_{13} & s^{}_{13} e^{-i\delta} \\ -s^{}_{12} c^{}_{23} -
c^{}_{12} s^{}_{13} s^{}_{23} e^{i\delta} & c^{}_{12} c^{}_{23} -
s^{}_{12} s^{}_{13} s^{}_{23} e^{i\delta} & c^{}_{13} s^{}_{23} \\
s^{}_{12} s^{}_{23} - c^{}_{12} s^{}_{13} c^{}_{23} e^{i\delta} &
-c^{}_{12} s^{}_{23} - s^{}_{12} s^{}_{13} c^{}_{23} e^{i\delta} &
c^{}_{13} c^{}_{23} \end{array} \right) P^{}_\nu \;,\label{pmns}
\end{equation}
where $c_{ij}=\cos\theta_{ij}$ and $s_{ij}=\sin\theta_{ij}$, $\theta_{12}$, $\theta_{23}$ and $\theta_{13}$ are the three mixing angles, 
$\delta_{CP}$ is the Dirac phase and the other two Majorana phases come in $P_{\nu}$ as
\begin{equation}\nonumber
P_{\nu}={\rm diag}(e^{i\rho},e^{i\sigma},1)\;.
\end{equation} 
The best-fit values and $3\sigma$ ranges of neutrino oscillation parameters taken from reference \cite{ri4}  are given in Table \ref{t12}.
One can reconstruct neutrino mass matrix once all the mixing parameters and three masses are known. Neutrino oscillation experiments probe mass 
squared differences ($\Delta m^2_{ij}=m_i^2-m_j^2$) and all the mixing parameters except Majorana phases. Results from neutrino oscillation experiments 
show two masses  are close to each other, while the third mass is comparatively far away than the other two ($\Delta m^2_{21}$ is of 
the order of $10^{-5}~{\rm eV}^2$ and $|\Delta m^2_{32}|$ is of the order of $10^{-3}~{\rm eV}^2$) and this results two possible  mass hierarchies, 
either $m_1<m_2<<m_3$ (normal hierarchy) or $m_3<<m_1<m_2$ (inverted hierarchy). There are ongoing neutrino oscillation experiments like NO$\nu$A and T2K etc.,
 which are expected to resolve mass hierarchy.  
 Beta decay experiments and cosmological bound on sum of neutrino masses ($\Sigma_i m_i$) give the absolute scale of neutrino masses while neutrino less 
double beta decay ($0\nu\beta\beta$) experiments can test Majorana nature of neutrinos. The tritium beta decay experiment KATRIN \cite{a1} shows absolute scale of neutrino mass is less than 0.35 eV and cosmological bound on    $\Sigma_i m_i$ from PLANCK data is 0.23 eV \cite{a2}.
 
 Initially neutrino oscillation experiments indicated the atmospheric  mixing angle, $\theta_{23}$ is maximal i.e., $\theta_{23}=\pi/4$ and 
reactor mixing angle $\theta_{13}$ is vanishingly small and  motivated by such anticipation many models for neutrino mixing were proposed such as 
Bimaximal mixing (BM) \cite{bm,bm1,bm2,bm3,bm4,bm5}, Tri-bimaximal mixing (TBM) \cite{tbm,tbm1,tbm2,tbm3,tbm4,tbm5,tbm6,tbm7,tbm8}, 
Golden ratio type-A (GRA), type-B (GRB) \cite{a4,a41} and Hexagonal mixing (HG), etc. 
All such models are based on some discrete symmetries such as $A_4$, $S_4$ \cite{a6,a7} etc and can be represented as
  \begin{equation}\nonumber
 \left(
\begin{array}{ccc}
\cos\theta_{12}&\sin\theta_{12}&0 \\

\frac{-\sin \theta_{12}}{\sqrt{2}}&\frac{\cos\theta_{12}}{\sqrt{2}}&\frac{-1}{\sqrt{2}}\\
\frac{-\sin\theta_{12}}{\sqrt{2}}&\frac{\cos\theta_{12}}{\sqrt{2}}&\frac{1}{\sqrt{2}}
\end{array}
\right)\;,
 \end{equation}
  where $\sin\theta_{12}$  takes the values $\frac{1}{\sqrt{2}}$, $\frac{1}{\sqrt{3}}$, $\frac{1}{\sqrt{2+r}}$ ($r=\frac{1+\sqrt{5}}{2}$ is the 
golden ratio), $\frac{\sqrt{3-r}}{2}$, and $\frac{1}{2}$  for BM, TBM, GRA, GRB and HG respectively. 
Recently  Daya Bay \cite{daya-bay, daya-bay1} RENO \cite{reno} and T2K \cite{t2k,t2k-result} experiments measured non-zero reactor mixing angle and hence,
 the above mentioned symmetry forms can't explain the experimental results.  But various studies show that these models can be modified suitably to 
accommodate the observed mixing angles by adding perturbations \cite{a5,a51,a52,a53,a54,a55,a56,a57,a58,a59,a510,a511,a512,a513,a514,
a515,a516,a517,a518,a519,a520,a521,a522,a523}.  
Among above mentioned symmetry forms TBM is of great interest because 
of its prediction to solar mixing angle, $\sin^2\theta_{12}=\frac{1}{3}$ against the experimental best fit value 0.323 and it can be explained on the basis 
of $A_4$ \cite{a6} symmetry, the smallest non abelian discrete symmetry with three dimensional irreducible representation.
The perturbations can be incorporated in various ways and one possible form for example is the  $Z_2\times Z_2$ symmetry  in neutrino sector 
and $Z_3$ symmetry in charged lepton sector. In this paper, we study a possible form of perturbation which modifies TBM to make it compatible with the 
recent experimental results. We also study the variation of electron neutrino mass 
($m_{\nu_e}$) and the 11 element of the Majorana neutrino mass matrix ($|M^{\nu}_{ee}|$), observables of $\beta$ decay and $0\nu\beta\beta$ decay experiments respectively with the lightest neutrino mass in order to verify the model.

The paper is  organized as follows. In section II, we will discuss briefly about the lepton mixing matrix and in section III we present 
the perturbation in the neutrino sector and its effect on the observables like mixing angles, $\delta_{CP}$, 
$m_{\nu_e}$ and $|M^{\nu}_{ee}|$. 
 We conclude our discussion 
in section IV.

\begin{table}[htb]
\begin{center}
\vspace*{0.1 true in}
\begin{tabular}{|c|c|c|}
\hline
 ~~Mixing Parameters~~ & ~~Best Fit values~~ & $~~ 3 \sigma $ Range~~  \\
\hline
$\sin^2 \theta_{12} $ &~ $0.323$ ~& ~$ 0.278 \to 0.375 $~\\

$\sin^2 \theta_{23}  $ (NH) &~ $0.567$ ~& ~$ 0.393 \to 0.643 $~\\

$\sin^2 \theta_{23}  $ (IH) &~ $0.573$ ~& ~$ 0.403 \to 0.640 $~\\

$\sin^2 \theta_{13} $ (NH) &~ $0.0226$ ~& ~$ 0.0190 \to 0.0262 $~\\

$\sin^2 \theta_{13} $ (IH) &~ $0.0229$ ~& ~$ 0.0193 \to 0.0265 $~\\

$\delta_{\rm CP}$ (NH) & ~$1.41 \pi$ & $ ~(0 \to 2 \pi)~ $\\
$\delta_{\rm CP}$ (IH) & ~$1.48 \pi$ & $ ~(0 \to 2 \pi)~ $\\
$\Delta m_{21}^2/ 10^{-5} {\rm eV}^2 $ & $ 7.60 $ & $ 7.11 \to 8.18 $ \\

$\Delta m_{31}^2/ 10^{-3} {\rm eV}^2 ({\rm NH}) $~ &~ $ 2.48
$ & $ 2.3 \to 2.65 $ \\

$\Delta m_{31}^2/ 10^{-3} {\rm eV}^2 ({\rm IH}) $ ~&~ $ -2.38 $ & $ -2.54 \to -2.20 $ \\

\hline
\end{tabular}
\end{center}
\caption{The best-fit values and the  $3\sigma$ ranges of the neutrino oscillation parameters from Ref. \cite{ri4}. }
\label{t12}
\end{table}


\section{The Lepton mixing matrix}
The lepton mixing matrix commonly known as PMNS matrix arises from the overlapping of the matrices that diagonalize charged lepton and neutrino mass matrices, hence PMNS matrix is given by
\begin{equation}
U_{PMNS}= U_l^{\dagger}U_{\nu}\;,
\end{equation} 
where $U_l$ and $U_{\nu}$ are the matrices which diagonalize charged lepton and neutrino mass matrices respectively. 
But it is always possible to work in a basis where charged lepton mass matrix is diagonal so that $U_l=I$ and $U_{PMNS}=U_{\nu}$. Hence, one can write 
\begin{equation}
U_{PMNS}=U_{\nu}\;,
\end{equation}
without loss of generality. So here we consider $U_l=I$ and $U_{\nu}$ as TBM in the leading order, and hence the PMNS matrix is given as  
\begin{equation}
U_{PMNS}=U_{TBM}=\left(
\begin{array}{ccc}
\frac{2}{\sqrt{6}}&\frac{1}{\sqrt{3}}&0 \\
\frac{-1}{\sqrt{6}}&\frac{1}{\sqrt{3}}&\frac{-1}{\sqrt{2}}\\
\frac{-1}{\sqrt{6}}&\frac{1}{\sqrt{3}}&\frac{1}{\sqrt{2}}
\end{array}
\right),\label{tbm}
\end{equation} 
Since TBM predicts $\theta_{13}=0$, it can't accommodate the recent observation of largish $\theta_{13}$ by the reactor experiments. So it has to be modified suitably for being in agreement with the experimental results. It is reasonable to assume that such modifications can come from perturbative corrections due to higher dimensional operators. We will discuss a possible form of perturbation in next section and show that it can consistently accommodate all the measured mixing angles.

\section{Perturbation in neutrino sector}

In this section we consider the deviations from TBM mixing angles  due to perturbation in the neutrino sector so that 
the obtained mixing angles satisfy experimental results. The perturbation is taken as a rotation in 23 plane followed by a rotation in 13 plane. 
Existence of Dirac CP phase is ensured by the complex phase in the 13 rotation matrix. Such a perturbation is quite reasonable, as
it will give correction to the atmospheric mixing angle $\theta_{23}$ which deviates from its maximal value and large
correction to the reactor mixing angle $\theta_{13}$.  With this perturbation the lepton mixing matrix will be of the form
\begin{equation}\label{e2}
U=U_{TBM} \cdot X\;,
\end{equation}
where $U_{TBM}$ is the TBM mixing matrix given in Eqn. (\ref{tbm}) and $X$
is the perturbation matrix given as
\begin{eqnarray}
X=\left(\begin{array}{ccc}
1 &0 & 0 \\
0 &c_{23}^{\prime}&s_{23}^{\prime}\\ 
0 &-s_{23}^{\prime}&c_{23}^{\prime}
\end{array}
\right)
\left(\begin{array}{ccc}
c_{13}^{\prime} &0 & s_{13}^{\prime}e^{-i\phi} \\
0 &1&0 \\
-s_{13}^{\prime}e^{i\phi} &0&c_{13}^{\prime}
\end{array}
\right)
=\left(\begin{array}{ccc}
c_{13}^{\prime} &0 & s_{13}^{\prime}e^{-i\phi} \\
-s_{23}^{\prime}s_{13}^{\prime}e^{i\phi} &c_{23}^{\prime}&s_{23}^{\prime}c_{13}^{\prime} \\
-c_{23}^{\prime}s_{13}^{\prime}e^{i\phi} &-s_{23}^{\prime}&c_{23}^{\prime}c_{13}^{\prime}
\end{array}
\right ) \;.
\end{eqnarray}
In general the leading order mixing matrix can receive corrections from both charged lepton and neutrino sector. For example, in Ref. \cite{ri3}, TBM mixing is realized based on $A_4$ symmetry which breaks to one of its subgroup $Z_3$ in the charged lepton sector while neutrino sector preserves $Z_2\times Z_2$ symmetry. They have shown that charged lepton and neutrino sector form a parallel world of flavour symmetry breaking and both the charged lepton and neutrino sectors receive corrections due to interaction between the sectors after symmetry breaking. But one can always go to charged lepton mass diagonal basis so that only neutrino sector contributes to lepton mixing.
We  obtained mixing angles and Jarlskog invariant in terms of elements of $U$ by equating it with PMNS matrix as
\begin{eqnarray}\label{e21}
\sin^2\theta_{12}&=&\frac{|U_{12}|^2}{1-|U_{13}|^2},~ ~~\sin^2\theta_{23}=\frac{|U_{23}|^2}{1-|U_{13}|^2},~~~\sin^2\theta_{13}=|U_{13}|^2 ,\nonumber\\ 
 J_{CP}&=& Im\left[U_{11}U_{22}U_{21}^*U_{12}^*\right]\;,
\end{eqnarray} 
where $U_{ij}$ is the $ij$ element of the lepton mixing matrix $U$.
Now comparing Eqns (\ref{e2}) and  (\ref{e21}), we obtain
\begin{eqnarray}
&&\sin^2\theta_{13}=\frac{1}{3}\left[ 2s_{13}^{\prime^2}+
2\sqrt{2} s_{23}^{\prime}c_{13}^{\prime}s_{13}^{\prime}\cos\phi +s_{23}^{\prime^2}c_{13}^{\prime^2}\right]\;,\\\label{e22}
&&\sin^2\theta_{12}=\frac{1-s_{23}^{\prime ^2}}{3-\left ( 2s_{13}^{\prime^2}+
2\sqrt{2} s_{23}^{\prime}c_{13}^{\prime}s_{13}^{\prime}\cos\phi +s_{23}^{\prime^2}c_{13}^{\prime^2}\right)}\;,\\ \label{e23}
&&\sin^2\theta_{23}=\frac{1}{2}-\frac{\sqrt{6}c_{23}^{\prime}c_{13}^{\prime}(s_{23}^{\prime}c_{13}^{\prime}-\frac{1}{\sqrt{2}}s_{13}^{\prime}\cos\phi)}
{3-\left( 2s_{13}^{\prime^2}+
2\sqrt{2} s_{23}^{\prime}c_{13}^{\prime}s_{13}^{\prime}\cos\phi +s_{23}^{\prime^2}c_{13}^{\prime^2}\right)}\;,\label{e24}
\end{eqnarray}
and
\begin{eqnarray}\label{e25}
J_{CP}&=&\frac{-1}{\sqrt{3}}\left (\frac{c_{23}^{\prime^2}}{3}-\frac{s_{23}^{\prime^2}}{2} \right )c_{23}^{\prime}c_{13}^{\prime}s_{13}^{\prime}\sin\phi\;.
\end{eqnarray}  
In standard parameterization the value of $J_{CP}$ is
\begin{equation}\label{e26}
J_{CP}=\frac{1}{8}\sin 2\theta_{12}\sin 2\theta_{23}\sin 2\theta_{13}\cos\theta_{13}\sin\delta_{CP}\;.
\end{equation}
Comparing equations (\ref{e25}) and (\ref{e26}) we obtain
\begin{eqnarray}
\sin\delta_{CP}=\frac{3(\frac{c_{23}^{\prime^2}}{3}-\frac{s_{23}^{\prime^2}}{2})c_{23}^{\prime}c_{13}^{\prime}s_{13}^{\prime}\sin\phi}
{\sqrt{X(\frac{2-X+s_{23}^{\prime^2}}{3-X})(\frac{1}{2}-\frac{Y}{3-X})(\frac{1}{2}+\frac{Y}{3-X})(3-X)}},\label{27}
\end{eqnarray}
where
\begin{eqnarray}
X&=&\left[ 2s_{13}^{\prime^2}+
2\sqrt{2} s_{23}^{\prime}c_{13}^{\prime}s_{13}^{\prime}\cos\phi +s_{23}^{\prime^2}c_{13}^{\prime^2}\right]\;,\nn\\
 Y&=&\sqrt{6}c_{23}^{\prime}c_{13}^{\prime}(s_{23}^{\prime}c_{13}^{\prime}-\frac{1}{\sqrt{2}}s_{13}^{\prime}\cos\phi)\;.
\end{eqnarray}
Next, we obtain the allowed parameter space  by varying these parameters
 $s_{23}^{\prime}$, $s_{13}^{\prime}$ and $\cos\phi$ in their allowed ranges i.e., between $-1$ to 1 and choosing those set of values for which the mixing angles fall within their $3\sigma$ ranges, which are shown in Fig. 1.  
\begin{figure}[!htb]
\includegraphics[width=7.0cm,height=5.0cm]{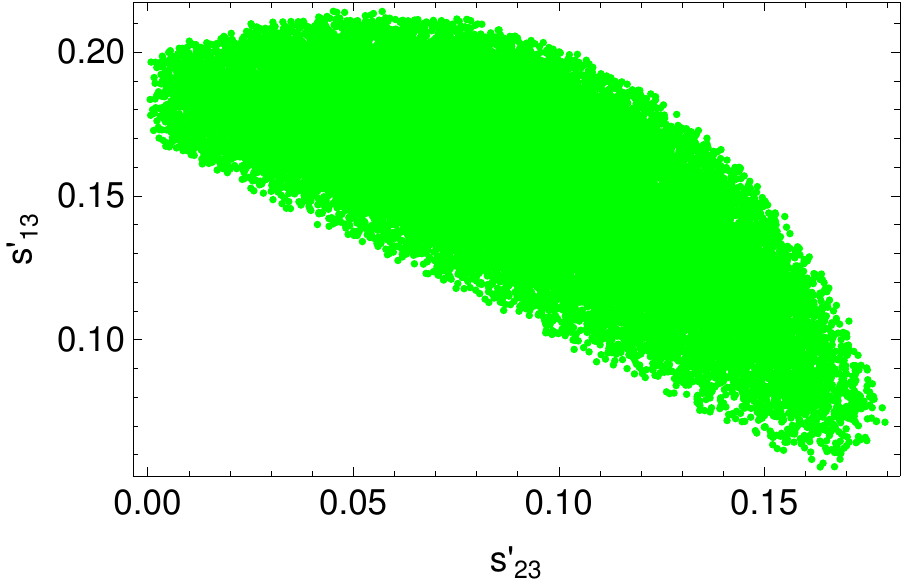}
\includegraphics[width=7.0cm,height=5.0cm]{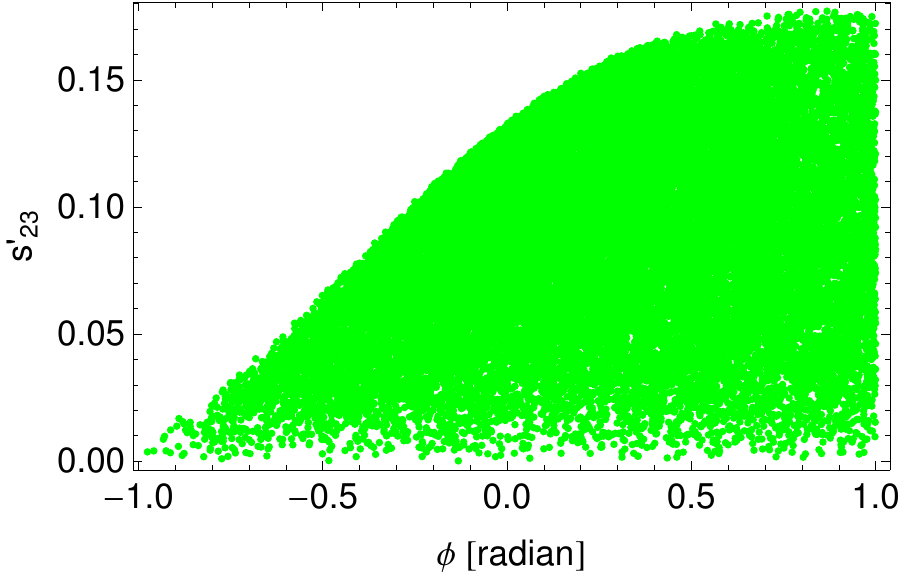}\\
\includegraphics[width=7.0cm,height=5.0cm]{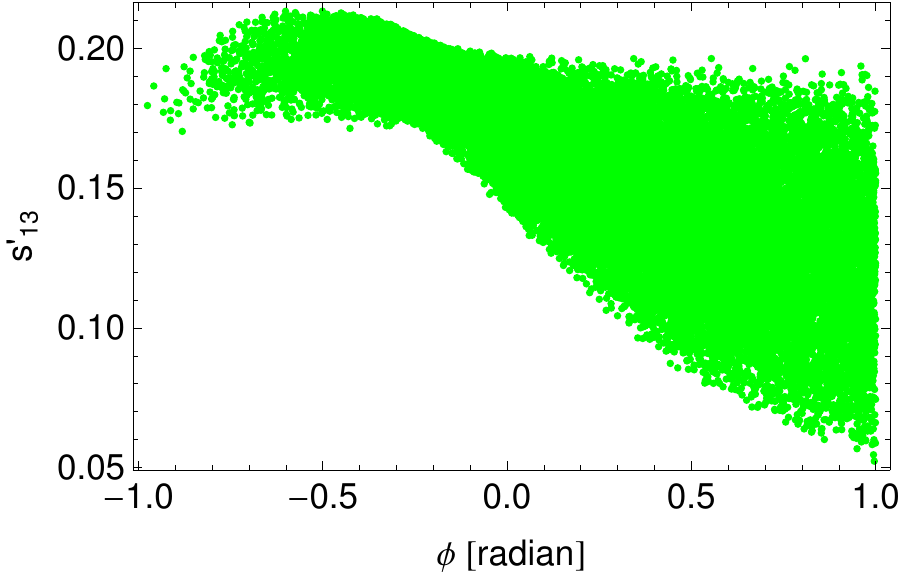}\\
\caption{Allowed parameter space in   $s_{13}^{\prime}-s_{23}^{\prime}$ , $s_{23}^{\prime}-\phi$ and $s_{13}^{\prime}-\phi$ planes compatible with the observed data.}
\end{figure}
Using the  allowed parameter space we show in Fig. 2, the correlation plots between the mixing angles, which are found to lie within their $3 \sigma$ allowed ranges. 
\begin{figure}[!htb]
\includegraphics[width=7.0cm,height=5cm]{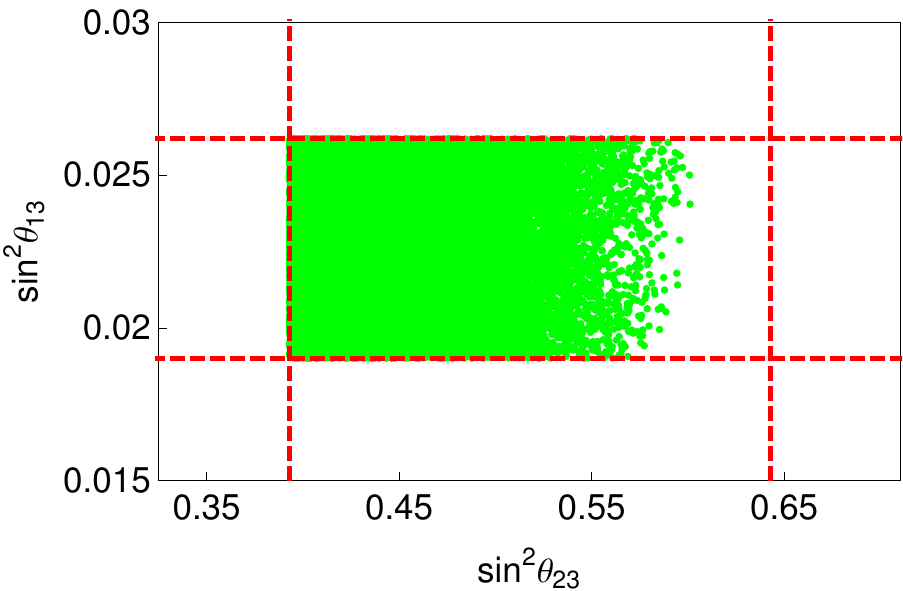}
\includegraphics[width=7.0cm,height=5cm]{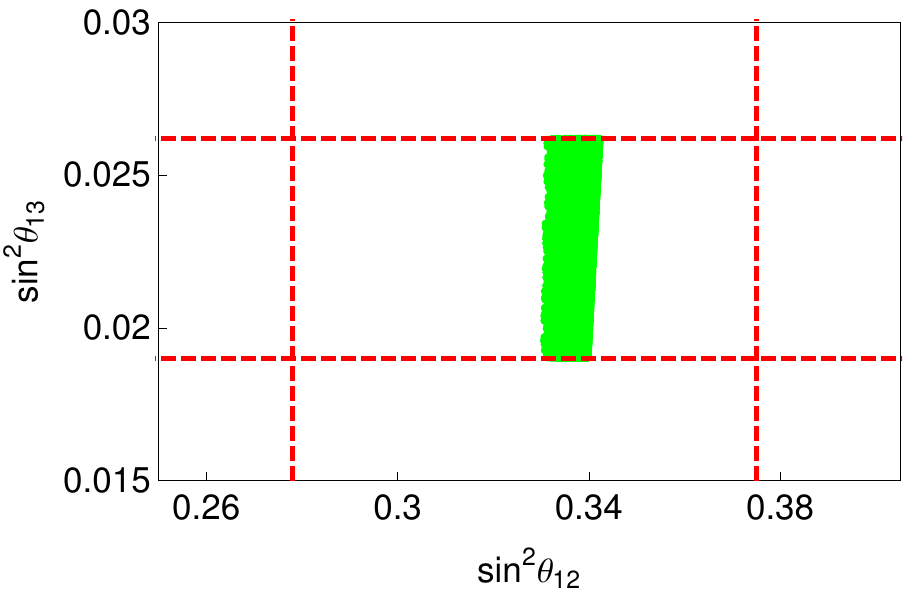}
\includegraphics[width=7.0cm,height=5cm]{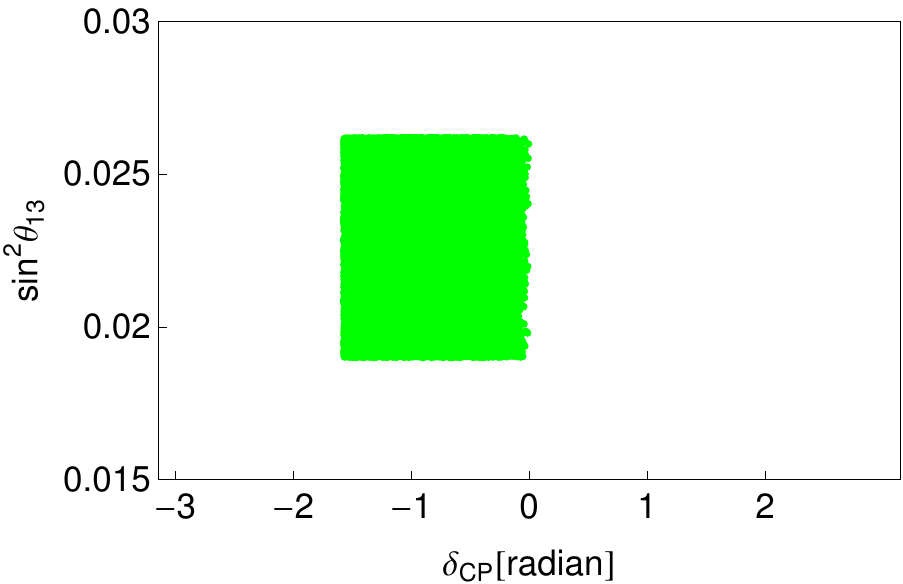}
\includegraphics[width=7.0cm,height=5cm]{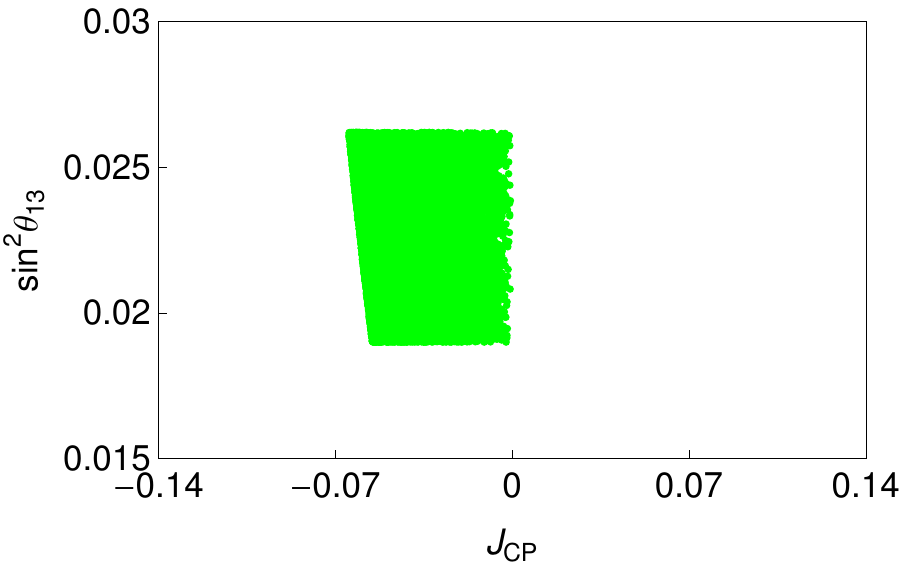}
\caption{Correlation plot between $\sin^2 \theta_{13}$ and $\sin^2 \theta_{23}$ (top left  panel), 
$\sin^2 \theta_{13}$ and $\sin^2 \theta_{12}$ (top right  panel),    $\delta_{CP}$ and
$\sin^2 \theta_{13}$  (bottom left  panel) and between $J_{CP}$ and $\sin^2 \theta_{13}$ (bottom right panel).}
\end{figure}

Neutrino oscillation experiments do not give any idea about the absolute mass of neutrinos as they only measure mass square differences. We will get the absolute scale of neutrino mass from Tritium beta decay experiments, which measure {electron neutrino mass} defined by
\begin{equation}
m_{\nu_e}=\Sigma_i|U_{1i}|^2m_i
\end{equation} 
where $i$ varies from 1 to 3 and $m_1$, $m_2$, and $m_3$ are light neutrino masses and $U_{1i}$'s are elements of first row of the lepton mixing matrix $ U$,
which are given by
\begin{eqnarray}
U_{11}&=& \frac{2}{\sqrt{6}}c_{13}^{\prime}-\frac{1}{\sqrt{3}}s_{23}^{\prime}s_{13}^{\prime}e^{i\phi}\;, \nonumber\\
U_{12} &=& \frac{1}{\sqrt{3}}c_{23}^{\prime}\;, \nonumber\\
U_{13}&=& \frac{2}{\sqrt{6}}s_{13}^{\prime}e^{-i\phi}+\frac{1}{\sqrt{3}}s_{23}^{\prime}c_{13}^{\prime}\; . \nonumber\\
\end{eqnarray}
We now proceed to  study the  variation of $m_{\nu_e}$ 
with lightest neutrino mass in the case of inverted  and normal hierarchy and the results are shown in the right panel of Fig 3. In our calculations we used the relations
\begin{eqnarray}
m_2 &=&\sqrt{\Delta m^2_{21}+m_1^2}\;, \nonumber\\
 m_3 &=&\sqrt{\Delta m_{31}^2+m_1^2}
\end{eqnarray}
for NH and
\begin{eqnarray}
m_1 &=&\sqrt{m_3^2-\Delta m^2_{31}}\;, \nonumber\\
 m_2 &=&\sqrt{m_3^2-\Delta m^2_{31}+\Delta m^2_{21}}
\end{eqnarray}
for IH
 and obtained upper limit on $m_1$ ($m_3$) as 0.071 eV (0.065 eV) in the case of normal (inverted) hierarchy  taking cosmological upper bound on $\Sigma_im_i$ as 0.23 eV \cite{a2}.
 
Neutrinos are very light compared to other fermions. The smallness of neutrino are addressed by different types of seesaw mechanisms such as, type-I, type-II and inverse seesaw. All those seesaw mechanisms treat neutrino as majorana particle. Hence they predict neutrinoless double beta decay,  a process in which two neutrons inside a nucleus 
convert to two protons without emitting neutrinos.
\begin{equation} \nonumber
(A,Z)\rightarrow(A,Z+2)+2e
\end{equation}  
The observation of neutrinoless double beta ($0\nu \beta \beta$) decay will be a consistency test for all those models. The half life of
 $0\nu \beta \beta$ decay  is proportional 
to $|M^{\nu}_{ee}|^2$ \cite{a8}, (1,1) element of neutrino mass matrix in flavor basis.
Several on going experiments like KamLAND-Ze \cite{r11}, EXO \cite{r12} and GERDA \cite{r13} to observe neutrino-less double beta decay 
 put upper bound on $|M^{\nu}_{ee}|$. The lowest upper bound on $|M^{\nu}_{ee}|$ is 0.22eV came from GERDA Phase-I data so only those mass models are valid which predicts $|M^{\nu}_{ee}|<0.22eV$. Hence we studied the variation of $|M^{\nu}_{ee}|$ with the lightest neutrino mass which is $m_1$ in the case of normal hierarchy and $m_3$ otherwise. We have calculated $|M^{\nu}_{ee}|$ using the relation
\begin{equation}
|M^{\nu}_{ee}|=|m_1 U_{11}^2+m_2U_{12}^2+m_3 U_{13}^2|.
\end{equation}
  The variation of $|M^{\nu}_{ee}|$ 
with lightest neutrino mass in the case of  normal and inverted hierarchies is shown in the left panel Fig 3. 

\begin{figure}[!htb]\label{fig4}
\includegraphics[width=6.5cm,height=5cm]{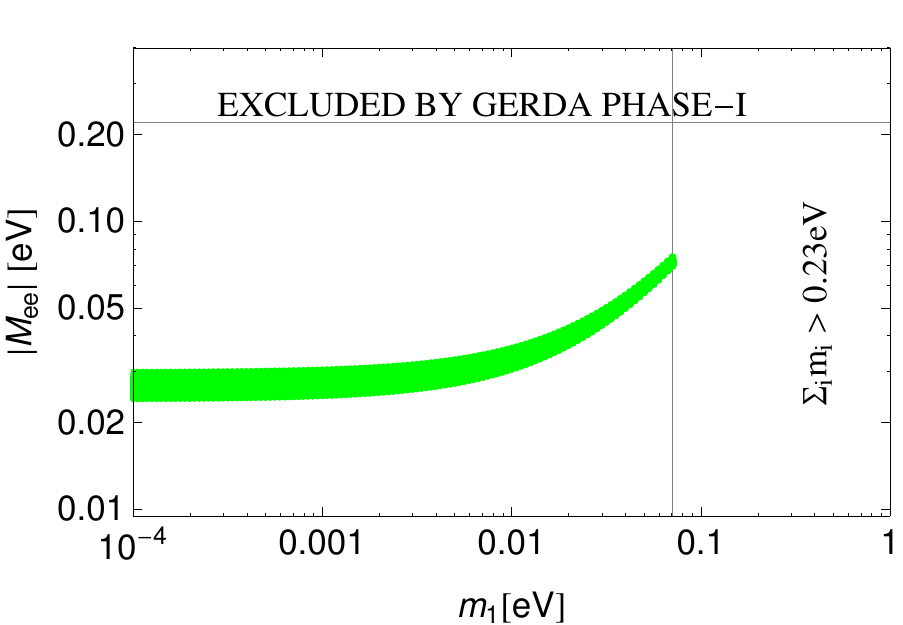}
\includegraphics[width=6.5cm,height=4.65cm]{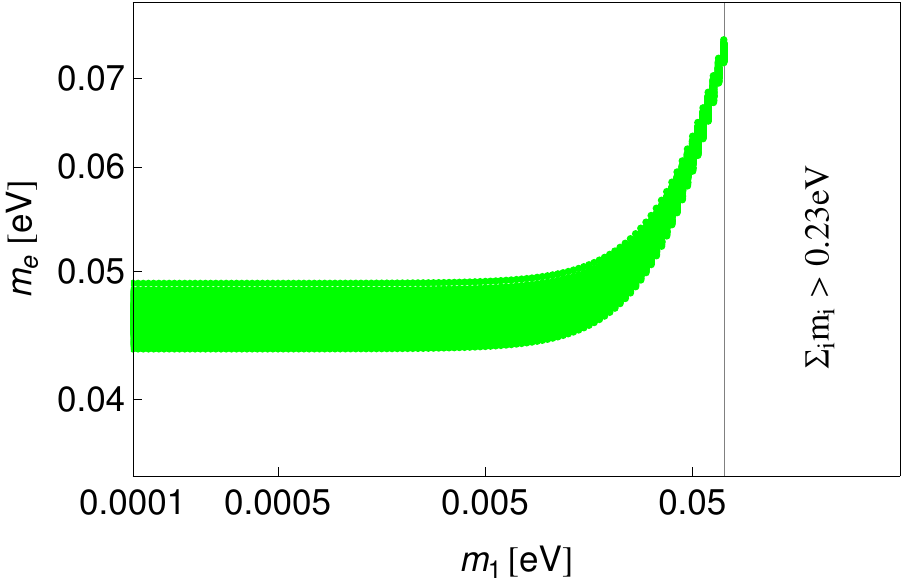}\\
\includegraphics[width=6.5cm,height=5cm]{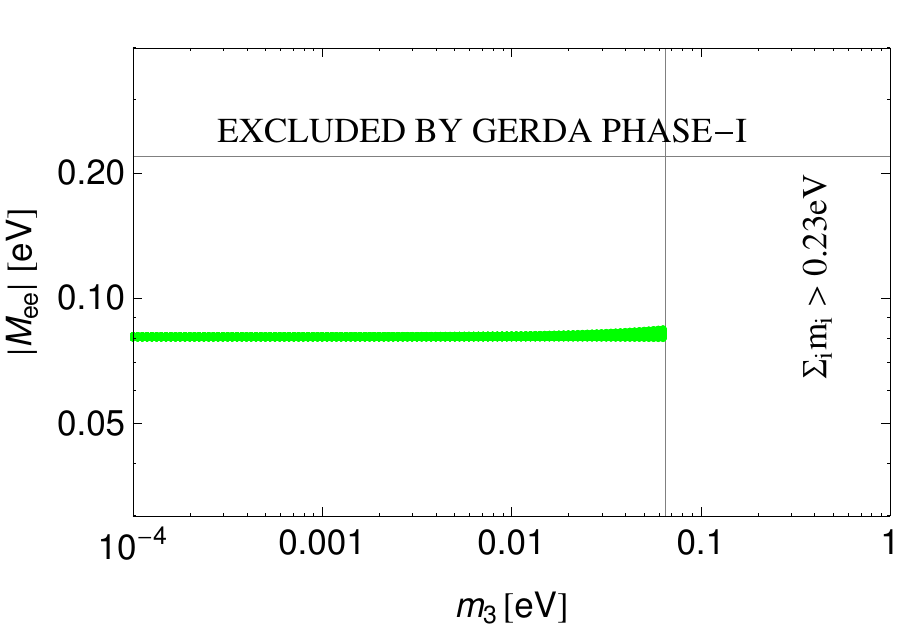}
\includegraphics[width=6.5cm,height=4.65cm]{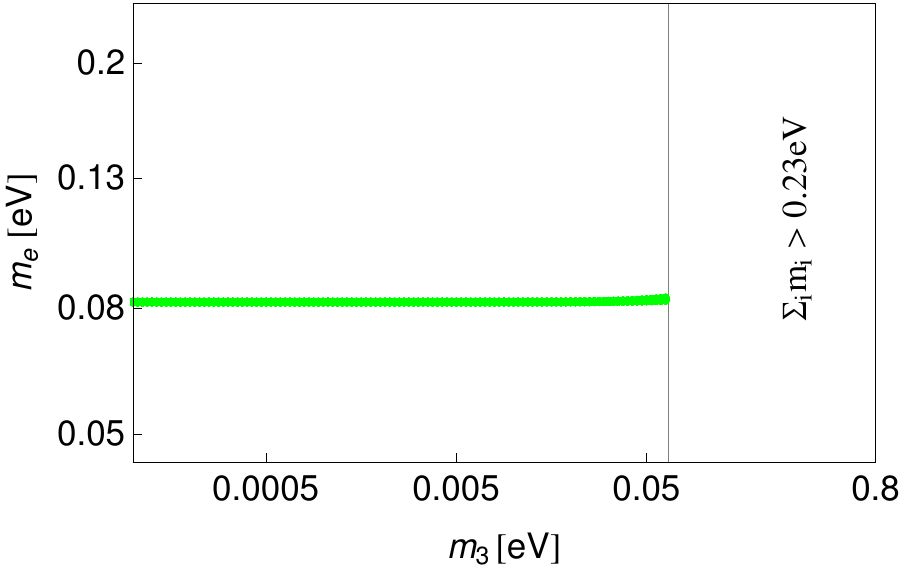}

\caption{variation of $|M^{\nu}_{ee}|$ with lightest neutrino mass $m_1$ ($m_3$) in the left panel and $m_{\nu_e}$ with $m_1$ ($m_3$) in the right panel for normal (inverted) hierarchy.}
\end{figure}

\section{Conclusions}
In this paper, we have studied deviation from Tribimaximal mixing (TBM) due to perturbation in neutrino sector in the form of combined rotation in 13 and 23 plane. It is found that such perturbation can explain the recent experimental results on neutrino mixing angles. We obtained the  parameter space for which mixing angles fall within their $3\sigma$ ranges and calculated possible values of Dirac CP phase ($\delta_{CP}$) and found that all values between 0 to $-\pi/2$ is possible for $\delta_{CP}$. We studied the variation of $|M^{\nu}_{ee}|$ with the lightest neutrino mass in the case of normal and inverted hierarchy and found that it falls bellow the upper bound (0.22 eV) for all values of the lightest neutrino mass below its upper bound, which we obtained as 0.071 eV for normal hierarchy and 0.065 for inverted hierarchy by taking cosmological upper bound on $\Sigma_im_i$ as 0.23 eV. We also studied the variation of electron neutrino mass, 
$m_{\nu_e}$ with lightest neutrino mass for normal and inverted mass hierarchies.


{\bf Acknowledgments}
SM  would like to thank University Grants Commission for financial support.

\end{document}